\documentclass[journal,twoside,web, a4paper, onecolumn]{ieeecolor}
\usepackage{tac}

\usepackage{cite}[sort,compress]
\usepackage{graphicx}
\usepackage{array}
\usepackage{hyperref}
\usepackage{arydshln}
\usepackage{float}
\usepackage{scrextend}
\usepackage{blindtext}
\usepackage{amsmath,amsfonts,amssymb}
\usepackage{tikz}
\usepackage{xspace}
\usepackage{xcolor}
\usepackage{mathrsfs,mathtools}
\usepackage{mathalfa}
\usepackage{euscript}
\usepackage{csquotes}
\usepackage{datetime}
\usepackage{arydshln}
\let\labelindent\relax
\usepackage{enumitem}

\newdateformat{monthyeardate}{\monthname[\THEMONTH], \THEYEAR}

\usepackage{tikz}
\usepackage{xspace}
\usepackage{xcolor}
\usepackage{amsthm}
\usepackage{amsmath,amssymb,amsfonts}
\usepackage{mathrsfs,mathtools}
\usepackage{mathalfa}
\usepackage{euscript}
\usepackage{csquotes}
\usepackage{enumitem}
\usepackage[normalem]{ulem}

\definecolor{mblue}{rgb}{0,0.4470,0.7410}
\definecolor{morange}{rgb}{0.8500,0.3250,0.0980}
\definecolor{myellow}{rgb}{0.9290,0.6940,0.1250}
\definecolor{mpurple}{rgb}{0.4940,0.1840,0.5560}
\definecolor{mgreen}{rgb}{0.4660,0.6740,0.1880}
\definecolor{mcyan}{rgb}{0.3010,0.7450,0.9330}
\definecolor{mred}{rgb}{0.6350,0.0780,0.1840}
\definecolor{mgreenblue}{rgb}{0.0,1.0,0.5}
\definecolor{parulablue}{rgb}{0.2431,0.1490,0.6588}
\definecolor{parulalblue}{RGB}{39,151,235}
\definecolor{parulagreen}{RGB}{129,204,89}
\definecolor{parulayellow}{RGB}{249,251,21}

\definecolor{cblue}{rgb}{0,0.9,1}
\definecolor{corange}{rgb}{1,0.7,0}
\definecolor{cgray}{rgb}{0.5,0.5,0.5}

\theoremstyle{definition}
\newtheorem{definition}{Definition}
\newtheorem{exmp}{Example}
\theoremstyle{plain}

\theoremstyle{remark}
\newtheorem{remark}{Remark}

\newenvironment{example}{\begin{exmp}}{\hfill$\blacktriangleleft$\end{exmp}}

\newcommand{\tss}[1]{\textsuperscript{#1}}

\newcommand{\comment}[1]{}

\newcounter{ass}

\newcommand{\mc}[1]{\mathcal{#1}}
\newcommand{\mf}[1]{\mathfrak{#1}}
\newcommand{\mr}[1]{\mathrm{#1}}
\newcommand{\mb}[1]{\mathbb{#1}}
\newcommand{\ms}[1]{\mathscr{#1}}

\newcommand{\mt}[1]{\mathtt{#1}}
\newcommand{\mbf}[1]{\mathbf{#1}}
\newcommand{\meu}[1]{\EuScript{#1}}
\DeclareFontFamily{U}{txcal}{\skewchar \font =45}
\DeclareFontShape{U}{txcal}{m}{n}{<-> txr-cal}{}
\DeclareMathAlphabet{\mathcalpxtx}{U}{txcal}{m}{n}

\newcommand{\col}{\mr{col}}

\newcommand{\kron}{\otimes} %

\newcommand{\svdots}{\raisebox{0pt}{$\scalebox{.75}{\vdots}$}}
\newcommand{\sddots}{\raisebox{0pt}{$\scalebox{.75}{$\ddots$}$}}

\newcommand{\dny}{n_\mr{y}}
\newcommand{\dnu}{n_\mr{u}}
\newcommand{\dnp}{n_\mr{p}}
\newcommand{\dnw}{n_\mr{w}}

\newcommand{\dnr}{n_\mr{r}}
\newcommand{\dna}{n_\mr{a}}
\newcommand{\dnb}{n_\mr{b}}

\newcommand{\deflen}[2]{%
    \expandafter\newlength\csname #1\endcsname
    \expandafter\setlength\csname #1\endcsname{#2}%
}

\title{Kernel-based multi-step predictors for data-driven analysis and control of nonlinear systems through the velocity form}

\author{%
    Chris Verhoek and Roland T{\'o}th%
    \thanks{This work has been supported by the European Union within the framework of the National Laboratory for Autonomous Systems (RRF-2.3.1-21-2022-00002).}
    \thanks{The authors are with the Control Systems group in the Dept. of Electrical Engineering at the Eindhoven University of Technology, The Netherlands. Roland T\'oth is also with the Institute for Computer Science and Control, Budapest, Hungary.}%
    \thanks{Corresponding author: C. Verhoek (\texttt{c.verhoek@tue.nl}).}%
}

\begin{document}
\maketitle

\thispagestyle{plain}
\pagestyle{plain}

\begin{abstract}
We propose kernel-based approaches for the construction of a single-step and multi-step predictor of the velocity form of nonlinear (NL) systems, which describes the time-difference dynamics of the corresponding NL system and admits a highly structured representation. The predictors in turn allow to formulate completely \emph{data-driven representations} of the velocity form. The kernel-based formulation that we derive, inherently respects the structured quasi-linear and specific time-dependent relationship of the velocity form. This results in an efficient multi-step predictor for the velocity form and hence for nonlinear systems. Moreover, by using the velocity form, our methods open the door for data-driven behavioral analysis and control of nonlinear systems with global stability and performance guarantees.
\end{abstract}

\section{Introduction}\label{s:intro}
In this technical note, we introduce a kernel-based method to establish single-step and multi-step predictors for nonlinear~(NL) systems. This in turn allows to formulate data-driven representations, which are useful for data-driven control and analysis of NL systems with global guarantees~\cite{verhoek2023direct}. The work presented in this note enables the extension and generalization of~\cite{VerhoekTothHaesaertKoch2021, verhoek2023direct}.

The key idea here is to formulate a data-driven characterization of the finite-horizon behavior of the velocity form of a NL system. {The velocity form is a specific representation that describes the time-difference dynamics of the NL system, see, e.g.,~\cite{Koelewijn2023}}. The advantage of using velocity forms of NL systems is that system properties, such as stability, of the velocity form, translate to equilibrium-\emph{independent} properties of the original NL systems. That means that with classic stability analysis of the velocity form, we can guarantee \emph{global} properties of the \emph{primal form} of the NL system~\cite{Koelewijn2023}. {Hence, a finite-horizon data-driven representation of the velocity form can then be used in all sorts of data-driven analysis and control applications, e.g., state-feedback control, predictive control or dissipativity analysis to achieve global guarantees in a data-driven setting.} Throughout, we refer to the original form of the NL system as the \emph{primal form}, while we use the \emph{velocity form} to refer to the velocity form of the NL system. {We want to emphasize that the velocity form is also nonlinear, but it has a specific quasi-linear structure. The matrix coefficient functions that constitute this quasi-linear relationship are dependent on the lagged version of the inputs and outputs of the primal form. We use kernel-based approaches for the construction of a non-parametric estimator of the velocity form. Instead of directly applying the method from~\cite{huang2023robust} for the general case, we exploit the quasi-linear and specific time-dependent relationship of the velocity form to establish a more efficient non-parametric estimator, i.e., data-driven representation, that can be use for control and analysis of NL systems with global guarantees.}

\begin{figure}[t]
    \centering
    \includegraphics[width=0.5\linewidth]{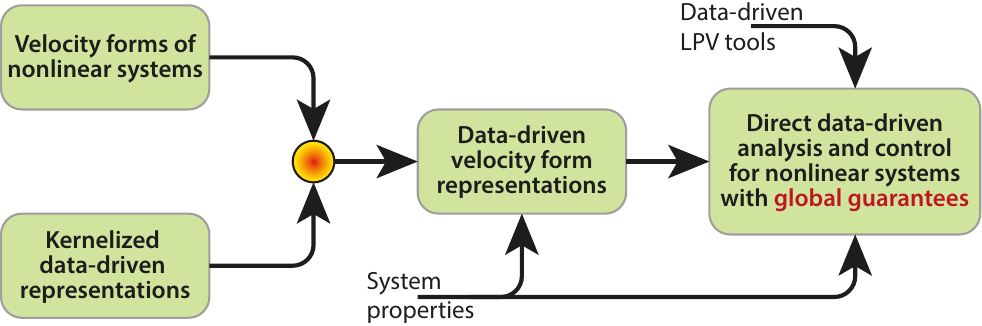}
    \caption{Visualization of the proposed approach and its utilization in data-driven analysis and controller synthesis for nonlinear systems.}\label{fig:scheme}
\end{figure}

This document introduces (i) a fully data-driven representation method for nonlinear systems in the behavioral setting without the need of a priori knowledge as in~\cite{verhoek2023direct} and~\cite[Sec.~VI.A]{verhoek2023dpcjournal}, and provides (ii) an extension and generalization of the data-driven methods discussed in~\cite{VerhoekTothHaesaertKoch2021}. The contribution is three-fold:
\begin{enumerate}
    \item We derive a method for a kernel-based formulation of the finite-horizon behavior of the velocity form. Recognising that this form can be considered as a linear parameter-varying (LPV) formulation, we can use the existing tools of LPV data-driven control to solve analysis and control problems for the underlying NL system. This proposed approach is illustrated in Fig.~\ref{fig:scheme} on the next page.
    \item The kernel-based formulation that we derive inherently respects the highly structured quasi-linear and specific time-dependent relationship of the velocity form.
    \item The kernel-based formulation that we derive is also useful for representing multi-step predictors for the \emph{primal} form of the NL system.
\end{enumerate}
We first discuss the velocity form that is directly derived from the primal form. This is followed by the derivation of the kernelized data-driven representations. We close this document with commenting on the usages of our proposed representations and also provide an example. 

\subsection*{Notation}
The set of positive integers is denoted {as} $\mathbb{N}$, while $\mathbb{R}$ denotes the set of real numbers. The $p$-norm of a vector $x\in\mathbb{R}^{n_\mathrm{x}}$ is denoted by $\lVert x\rVert_p$ {and} the Moore-Penrose (right) pseudo-inverse {of a matrix} is denoted by $\dagger$. {The} Kronecker product of two matrices $A$ and $B$ {is $A\kron B$}. The identity matrix of size $n$ is denoted as $I_n$ and $0_{n\times m}$ denotes the $n\times m$ zero-matrix. For $\mb{A}$ and $\mb{B}$, $\mb{B}^\mb{A}$ indicates the collection of all maps from $\mb{A}$ to $\mb{B}$.
The notation  $A \succ 0$ and $A\prec 0$ ($A \succeq0 $ and $A \preceq 0$) stands for positive/negative (semi) definiteness of $A\in\mathbb{R}^{n\times n}$. A block-matrix of the form $\begin{bsmallmatrix}A & 0 \\ 0 & B\end{bsmallmatrix}$ is denoted in short as $\mr{blkdiag}(A,B)$.
The value of a signal $w:\mb{Z}\to\mb{R}^{\dnw}$ at time step $k$ is denoted as $w(k)\in\mb{R}^{\dnw}$. Considering the time interval $\mb{T}\subseteq\mb{Z}$, we write $w_\mb{T}$ as the truncation of $w$ to $\mb{T}$, e.g., for $\mb{T}:=[1,N]$ we have $w_{[1,N]}=(w(1),\dots,w(N))$. For the sequence of a collection of signals, e.g., $((w_1(1),w_2(1)),\dots,  (w_1(N),w_2(N)))$ we use the short-hand notation $(w_1(k),w_2(k))_{k=1}^N$. 
For the sequence $w_{[1,N]}$, the associated Hankel matrix of depth $L$ is given as
\begin{equation}\label{eq:notation:hankel}
\mc{H}_L(w_{[1,N]})=\begin{bmatrix}w(1) & w(2) & \dots & w({N-L+1})\\
w(2) & w(3) & \dots & w(N-L+2)\\
\vdots & \vdots & \ddots & \vdots \\
w({L}) & w({L+1}) & \dots & w({N})
\end{bmatrix}.
\end{equation}
If a function $f$ is an element of $\mc{C}_n$, it means that it is $n$ times continuously differentiable.

\section{Velocity forms of nonlinear input-output representations}

We consider causal nonlinear (NL) systems defined in terms of input-output (IO) representations with finite recurrence, i.e., the current output is  nonlinearly dependent on the current input and a finite number of past input and output samples. There are multiple forms that can be considered. We will work out the most general form and a more structured form, which will be useful later in this document.
\subsection{General form}\label{ss:genvel}
Based on the aforementioned setting, we consider systems whose dynamics can be represented in terms of the general NL-IO representation:
\begin{equation}\label{eq:io:gen}
    y(k) = f(y(k-1), \dots, y(k-\dna), u(k), \dots, u(k-\dnb)),
\end{equation}
with $y(k)\in\mb{R}^{\dny}$, $u(k)\in\mb{R}^{\dnu}$, $f\in\mc{C}_1$, and $\dna>0, \dnb\geq0$. {We will refer to~\eqref{eq:io:gen} as the \emph{primal form} of the nonlinear system.}
To obtain the velocity form of the system, we will analyze the time-difference dynamics of this IO form, i.e., the dynamics involving $\Delta u(k):= u(k)-u(k-1)$ and $\Delta y(k):= y(k)-y(k-1)$:
\begin{multline}\label{eq:differenceformgeneral}
    y(k)-y(k-1) = f(y(k-1), \dots, y(k-\dna), u(k), \dots, u(k-\dnb)) \\- f(y(k-2), \dots, y(k-\dna-1), u(k-1), \dots, u(k-\dnb-1)). 
\end{multline}
Let $\nu_i(k):=y(k-i)$ and $\mu_i(k):=u(k-i)$, and define for 
\[\xi(k):=\begin{bmatrix}\nu_1(k)^\top & \cdots &\nu_{\dna}(k)^\top & \mu_0(k)^\top &  \cdots & \mu_{\dnb}(k)^\top \end{bmatrix}^\top\]
the `wrapper' function $\bar{f}(\xi(k)) := f(\xi_1(k), \dots, \xi_{\dna+\dnb+1}(k))$. 
Applying the \emph{Fundamental Theorem of Calculus} (FTC) on $\bar{f}$ w.r.t. $\xi$ gives:
\begin{align}
    \Delta y(k) & = \left(\int_0^1\frac{\partial \bar f}{\partial \xi}(\bar{\xi}(k,\lambda))\mr{d}\lambda\right) \left(\xi(k)-\xi(k-1)\right), \quad \bar{\xi}(k,\lambda):= \xi(k-1) + \lambda(\xi(k)-\xi(k-1)), \\ 
     & = \sum_{i=1}^{\dna}\underbrace{\left(\int_0^1\frac{\partial f}{\partial \nu_i}(\bar{\xi}(k,\lambda))\mr{d}\lambda\right)}_{\mf{a}_i(y(k-1),\dots, u(k-\dnb-1))} \Delta y(k-i) + \sum_{j=0}^{\dnb}\underbrace{\left(\int_0^1\frac{\partial f}{\partial \mu_j}(\bar{\xi}(k,\lambda))\mr{d}\lambda\right)}_{\mf{b}_j(y(k-1),\dots, u(k-\dnb-1))} \Delta u(k-j),\label{eq:velformIntegrals}
\end{align}
which results, with more compact notation, in the IO form:
\begin{equation}\label{eq:nlvelioform}
    \Delta y(k) = \sum_{i=1}^{\dna}\mf{a}_i(w_k)\Delta y(k-i) + \sum_{j=0}^{\dnb}\mf{b}_j(w_k)\Delta u(k-j),
\end{equation}
{where we collected the lagged versions of $u$ and $y$ in the signal $w_k$ that we define as
\begin{equation}\label{eq:defwk}
    w_k := \mr{col}\big(y(k-1), \dots, y(k-\dna-1), u(k), \dots, u(k-\dnb-1)\big).
\end{equation}
This is called the \emph{velocity form} of the NL-IO representation~\eqref{eq:io:gen}.}
This IO representation of the velocity form of~\eqref{eq:io:gen} is linear in the $\Delta$-signals and this linear, dynamic relationship can be seen to be varying along the signals $y(k-1), \dots,y(k-\dna-1),u(k),\dots,u(k-\dnb-1)$. Later in this note, we will propose a method for the formulation of a finite-horizon data-driven representation of~\eqref{eq:nlvelioform}, where the functional relationship of the matrix functions $\mf{a}_i(\cdot), \mf{b}_j(\cdot)$ are captured using kernel-based methods. Note that next to the velocity form, we can also use the FTC for factorizing the primal form to a similar quasi-linear representation of the system (see Appendix~\ref{app:primalform} for details). 

\subsection{Structured forms}\label{ss:shiftvel}
One can consider various structured forms of~\eqref{eq:io:gen}, for which the complexity of the matrix functions $\mf{a}_i(\cdot), \mf{b}_j(\cdot)$ reduces. We consider the case where the number of inputs of these matrix functions is reduced by considering a so-called \emph{shifted} structure. That is, a structure that does not have cross terms between $(y(k-i),u(k-i))$ and $(y(k-j),u(k-j))$. For notational simplicity, we do not consider direct feed-through for this form. This results in the following NL-IO realization:
\begin{equation}\label{eq:io:str1}
    y(k) = \sum_{i=1}^{n} f_i(y(k-i),u(k-i)).
\end{equation}
By analyzing the time-difference dynamics and application of the FTC, we arrive at the following compact IO representation of the velocity form of~\eqref{eq:io:str1}:
\begin{equation}\label{eq:nlvelioform-shifted}
    \Delta y(k) = \sum_{i=1}^{n}\Big(\bar{\mf{a}}_i\!\left(w_{k-i}\right)\Delta y(k-i) + \bar{\mf{b}}_i\!\left(w_{k-i}\right)\Delta u(k-i) \Big), 
\end{equation}
where $w_k:= \mr{col}(y(k),y(k-1),u(k),u(k-1))$. The complete derivation can be found in Appendix~\ref{app:deriveshiftedvel}.
Hence, now we have a linear-in-$\Delta$, \emph{shifted} IO representation of the velocity form. The IO representation has a linear relationship w.r.t. the $\Delta$-signals, while this relationship is varying in a time-{shifted} manner along $y(k-i),y(k-i-1),u(k-i),u(k-i-1)$, i.e., the time dependence of the parametrization is linked to the corresponding $\Delta$-signal. Compared to the general velocity form of~\eqref{eq:io:gen}, the matrix functions $\bar{\mf{a}}_i(\cdot), \bar{\mf{b}}_j(\cdot)$ are only dependent on $y(k-i),y(k-i-1),u(k-i),u(k-i-1)$, instead of $y(k-1),\dots, y(k-\dna),u(k),\dots, u(k-\dnb)$, which reduces the complexity.

\subsection{Computation of the matrix functions $\mf{a}_i(\cdot), \mf{b}_j(\cdot)$}
For certain functions $f_{(i)}$, we can explicitly compute the integrals in~\eqref{eq:velformIntegrals} to compute velocity form, e.g., for polynomial $f$. Consider the following example and for simplicity introduce the notation $u(k-1)=u_-$. Then:
\begin{multline*}
    f(u) = u+u^2+u^3 + \dots + u^r \ \overset{\bar{u}(\lambda) = u_-+\lambda(u-u_-)}{\longrightarrow} \ \int_0^1\frac{\partial f}{\partial u}(\bar{u}(\lambda))\mr{d}\lambda = \int_0^1( 1+ 2\bar{u}(\lambda) + 3\bar{u}(\lambda)^2 + \dots + r\bar{u}(\lambda)^{r-1},  \\
    = \int_0^1\mr{d}\lambda + 2 \int_0^1(u_- + \lambda(u-u_-))\mr{d}\lambda +\dots + r \int_0^1(u_- + \lambda(u-u_-))^{r-1}\mr{d}\lambda \\ = 1 + u + u_- + \dots + \frac{u^i-u_-^i}{u-u_-} + \dots + \frac{u^r-u_-^r}{u-u_-},
\end{multline*}
giving:
\[ \Delta y(k) = \left(1 + u(k) + u(k-1) + \dots + \frac{u(k)^i-u(k-1)^i}{u(k)-u(k-1)} + \dots + \frac{u(k)^r-u(k-1)^r}{u(k)-u(k-1)}\right)\Delta u(k). \] 
Note that the above expression is well defined for $u(k)=u(k-1)$ as in that case $\lim_{h\to0}\frac{x^r-(x+h)^r}{x-(x+h)}=r x^{r-1}$, i.e., the limit is equal to the partial derivative of the element. In the general case, we can use either symbolic integration or numerical integration, such as the Simpson rule for the computation of $\mf{a}_i(\cdot), \mf{b}_j(\cdot)$, see~\cite{hoekstra2023computationally} for further details.

\subsection{Basis function representation of the IO forms}
If the system is not know, the velocity form~\eqref{eq:nlvelioform} requires the estimation of the matrix functions $\mf{a}_i(\cdot)$ and $\mf{b}_j(\cdot)$ from data. To facilitate this, the matrix functions $\mf{a}_i(\cdot), \mf{b}_j(\cdot)$ can be parametrized in terms of a linear combination set of basis functions. That is, we can write the underlying dependency structure as a (possibly infinite) sum of basis functions $\psi_s:\mb{R}^{\dnw}\to\mb{R}$, $s = 0, \dots, {n_\psi}$, i.e., 
\begin{subequations}\label{eq:basisexpansion}
\begin{equation}\label{eq:basisexpansion:a}
    \mf{a}_i(w_k) = \sum_{s=0}^{n_\psi}a_{i,s}\psi_s(w_{k}), \quad \forall i = 1,\dots,\dna,
\end{equation}
and
\begin{equation}\label{eq:basisexpansion:b}
    \mf{b}_j(w_k) =\sum_{s=0}^{n_\psi}b_{j,s}\psi_s(w_{k}), \quad \forall j = 0,\dots,\dnb,
\end{equation}
\end{subequations}
where $a_{i,s}\in\mb{R}^{\dny\times\dny}$, $b_{j,s}\in\mb{R}^{\dny\times\dnu}$ and we assume $\psi_0(w_k)=1$ for any $w_k$ without loss of generality. Let us write $\psi(w_k)$ for the vector $\begin{bmatrix} \psi_1(w_k) & \cdots & \psi_{n_\psi}(w_k) \end{bmatrix}^\top$ (i.e., without $\psi_0(w_k)=1$).

\begin{remark} If $\mf{a}_i(\cdot), \mf{b}_j(\cdot)$ are the coefficient matrix functions of the shifted representation, we have that $w_k$ shifts along in the basis function expansion. This gives the variation of~\eqref{eq:basisexpansion:a} as $\bar{\mf{a}}_i(w_{k-i}) = \sum_{s=0}^{n_\psi}\bar{a}_{i,s}\psi_s(w_{k-i})$, similarly for $\bar{\mf{b}}_i$.
\end{remark}
\begin{remark}
    Suppose we \emph{know} the \emph{finite} set of basis functions $\psi$ that realizes~\eqref{eq:basisexpansion}, then the assumption that is taken in~\cite[Assump.~1]{verhoek2023direct} holds and we can apply the results of~\cite{verhoek2023direct} directly to achieve direct data-driven control of general nonlinear systems. In the next section, we solve the problem of overcoming the need of knowing this (possibly infinite) set of basis functions and thus to \emph{avoid}~\cite[Assump.~1]{verhoek2023direct}. 
\end{remark}

\section{Regressor form of the multi-step predictor}\label{s:regressor}
The key aspect of our main result in this note is the exploitation of the structure that is present in the NL-IO representation of the velocity form. Therefore, we first derive a regressor form for a multi-step predictor, which gives us insight in the structure of the problem. This is followed by applying a \emph{Support Vector Machine} (SVM) approach to the problem to arrive at a kernelized {multi-step predictor, that we can use for the formulation of a data-driven representation of the velocity form.}

\subsection{Exploring the structure of the multi-step predictor}

Reconsider the IO representation for the velocity form~\eqref{eq:nlvelioform} of~\eqref{eq:io:gen}:
\begin{equation*}
    \Delta y(k) = \sum_{i=1}^{\dna}\mf{a}_i(w_{k})\Delta y(k-i) + \sum_{j=0}^{\dnb}\mf{b}_j(w_{k})\Delta u(k-j),
\end{equation*}
and recall $w_k=\mr{col}\big(y(k-1), \dots, y(k-\dna-1), u(k), \dots, u(k-\dnb-1)\big)$. Let $\dnr=\max\{\dna,\dnb\}$. For the construction of a data-driven representation, we consider the basis function representation of the IO forms, i.e., $\mf{a}_i$ and $\mf{b}_j$ are written as in~\eqref{eq:basisexpansion}.
We want to obtain a multi-step predictor that has the following form:
\begin{equation}\label{eq:multisteppredictorformdesired}
    \Delta y_{[1,L]} = \ms{T}(w_{[1-\dnr,L]})\begin{bmatrix} \Delta y_{[1-\dnr, 0]} \\ \Delta u_{[1-\dnr, 0]} \\ \Delta u_{[1,L]}\end{bmatrix}\begin{matrix} \left.\vphantom{\begin{matrix}y_{[1-\dnr, 0]} \\ u_{[1-\dnr, 0]}\end{matrix}}\right\}\Delta\varphi_0 \\ \vphantom{u_{[1,L]}} \end{matrix}
\end{equation}
with the \emph{regressor} $\begin{bmatrix} \Delta\varphi_0^\top & \Delta u_{[1,L]}^\top\end{bmatrix}^\top$. We can expose the structure of $\ms{T}(w_{[1-\dnr,L]})$ by means of an example:
\begin{example}
For the sake of illustration, we consider a simple system with $\dna=1$, $\dnb=1$ and $\dnu=\dny=1$ (SISO case). To improve the clarity of the presentation, we take a few simplifying notations to clarify the presentation: (i) we disregard the $\Delta$ prefixes for the (collection of) signals $\Delta y, \Delta u,\Delta\varphi_0$ to avoid notational clutter. (ii) with a slight abuse of notation, we write $\psi(k):=\psi_1(w_k)$. The basis functions based expansion~\eqref{eq:basisexpansion} of the coefficient functions for this simple system is as follows:
\[ y(k) = a_{1}y(k-1) + a_2 \psi(k) y(k-1) + b_{1}u(k-1) + b_2 \psi(k) u(k-1) \]
that we can conveniently write as:
\begin{equation}
    y(k) = \underbrace{\begin{bmatrix}a_{1} & a_2 & b_1 & b_2 \end{bmatrix}}_{\theta_{k}} \begin{bmatrix} y(k-1) \\ \psi(k) y(k-1) \\u(k-1) \\\psi(k) u(k-1) \end{bmatrix} = \theta_k \underbrace{\begin{bmatrix}\begin{bsmallmatrix} 1 \\ \psi(k) \end{bsmallmatrix}\kron I_{\dny} & 0 \\ 0 & \begin{bsmallmatrix} 1 \\ \psi(k) \end{bsmallmatrix}\kron I_{\dnu}\end{bmatrix}}_{{\Psi}(w_{k})} \begin{bmatrix} y(k-1)\\u(k-1) \end{bmatrix}.
\end{equation}
Writing out the first three time-steps reveals the enormous amount of structure we can exploit later:
\begin{subequations}
\begingroup\allowdisplaybreaks
\begin{align}
    y(1) & = \theta_1 \underbrace{\begin{bmatrix}\begin{bsmallmatrix} 1 \\ \psi(1) \end{bsmallmatrix}\kron I_{\dny} & 0 \\ 0 & \begin{bsmallmatrix} 1 \\ \psi(1) \end{bsmallmatrix}\kron I_{\dnu}\end{bmatrix}}_{{\Psi}(w_{1})} \begin{bmatrix} y(0)\\u(0) \end{bmatrix} = \theta_1{\Psi}(w_{1})\varphi_0, \\
    y(2) & = \begin{bmatrix}a_{1} & a_2 & b_1 & b_2 \end{bmatrix}\begin{bmatrix} y(1) \\ \psi(2) y(1) \\u(1) \\\psi(2)  u(1) \end{bmatrix} = a_1 \theta_1 \begin{bmatrix} y(0) \\ \psi(1) y(0) \\u(0) \\\psi(1) u(0) \end{bmatrix} + a_2 \psi(2)\theta_1 \begin{bmatrix} y(0) \\ \psi(1) y(0) \\u(0) \\\psi(1) u(0) \end{bmatrix} + \begin{bmatrix} b_1 & b_2 \end{bmatrix}\begin{bmatrix} u(1) \\\psi(2) u(1) \end{bmatrix} \notag \\
    & = \underbrace{\begin{bmatrix} a_1\theta_1 & a_2 \theta_1 & b_1 & b_2 \end{bmatrix}}_{\theta_2} \begin{bmatrix}
        \begin{bmatrix} 1 \\ \psi(2) \end{bmatrix}\kron I_{(1+n_\psi)(\dnu+\dny)} & 0 \\ 0 & \begin{bmatrix} 1 \\ \psi(2) \end{bmatrix}\kron I_{\dnu} \end{bmatrix}\begin{bmatrix} y(0) \\ \psi(1) y(0) \\u(0) \\\psi(1) u(0) \\ u(1) \end{bmatrix} \notag\\
    & = \theta_2 \underbrace{\begin{bmatrix}
        \begin{bmatrix} 1 \\ \psi(2) \end{bmatrix}\kron I_{(1+n_\psi)(\dnu+\dny)} & 0 \\ 0 & \begin{bmatrix} 1 \\ \psi(2) \end{bmatrix}\kron I_{\dnu} \end{bmatrix}\begin{bmatrix} \Psi(w_{1}) & 0 \\ 0 & I_{\dnu}\end{bmatrix}}_{\Psi(w_{[1,2]})}\begin{bmatrix} y(0)\\u(0)\\ u(1)\end{bmatrix} = \theta_2\Psi(w_{[1,2]})\begin{bmatrix}\varphi_0 \\ u(1) \end{bmatrix},\\
    y(3) & = \begin{bmatrix}a_{1} & a_2 & b_1 & b_2 \end{bmatrix}\begin{bmatrix} y(2) \\ \psi(3) y(2) \\u(2) \\\psi(3) u(2) \end{bmatrix} = a_1 \theta_2 \begin{bmatrix} y(0) \\ \psi(1) y(0) \\u(0) \\\psi(1) u(0) \\ \psi(2)y(0) \\ \psi(2)\psi(1) y(0) \\\psi(2)u(0) \\\psi(2)\psi(1) u(0) \\ u(1) \\ \psi(2)u(1) \end{bmatrix} + a_2 \psi(3)\theta_2 \begin{bmatrix} y(0) \\ \psi(1) y(0) \\u(0) \\\psi(1) u(0) \\ \psi(2)y(0) \\ \psi(2)\psi(1) y(0) \\\psi(2)u(0) \\\psi(2)\psi(1) u(0) \\ u(1) \\ \psi(2)u(1) \end{bmatrix} + \begin{bmatrix} b_1 & b_2 \end{bmatrix}\begin{bmatrix} u(2) \\\psi(3) u(2) \end{bmatrix} \notag \\
    & = \underbrace{\begin{bmatrix} a_1\theta_2 & a_2 \theta_2 & b_1 & b_2 \end{bmatrix}}_{\theta_3} \begin{bmatrix}
        \begin{bmatrix} 1 \\ \psi(3) \end{bmatrix}\kron I_{(n_\psi+1)(\dnu+(1+n_\psi)(\dnu+\dny))} & 0 \\ 0 & \begin{bmatrix} 1 \\ \psi(3) \end{bmatrix}\kron I_{\dnu} \end{bmatrix}\left.\begin{bmatrix} y(0) \\ \psi(1) y(0) \\u(0) \\\psi(1) u(0) \\ \psi(2)y(0) \\ \psi(2)\psi(1) y(0) \\\psi(2)u(0) \\\psi(2)\psi(1) u(0) \\ u(1) \\ \psi(2)u(1) \\ u(2) \end{bmatrix}\right\}\begin{bmatrix}\Psi(w_{[1,2]})\begin{bmatrix}\varphi_0\\u(1)\end{bmatrix}\\u(2)\end{bmatrix} \notag\\
    & = \theta_3\underbrace{\begin{bmatrix}
        \begin{bmatrix} 1 \\ \psi(3) \end{bmatrix}\kron I_{(n_\psi+1)(\dnu+(1+n_\psi)(\dnu+\dny))} & 0 \\ 0 & \begin{bmatrix} 1 \\ \psi(3) \end{bmatrix}\kron I_{\dnu} \end{bmatrix}\begin{bmatrix} \Psi(w_{[1,2]}) & 0 \\ 0 & I_{\dnu}\end{bmatrix}}_{\Psi(w_{[1,3]})}\begin{bmatrix} \varphi_0\\ u(1)\\u(2)\end{bmatrix}.
\end{align}
\endgroup
Hence, for a general time-step, we have for $i\geq1$:
\begin{equation}\label{eq:exmp:general}
    y(i) = \underbrace{\begin{bmatrix} a_1\theta_{i-1} & a_2 \theta_{i-1} & b_1 & b_2 \end{bmatrix}}_{\theta_i} \underbrace{\begin{bmatrix}
        \begin{bmatrix} 1 \\ \psi(i) \end{bmatrix}\kron I_{n_{\Psi_i}} & 0 \\ 0 & \begin{bmatrix} 1 \\ \psi(i) \end{bmatrix}\kron I_{\dnu} \end{bmatrix}\begin{bmatrix} \Psi(w_{[1,i-1]}) & 0 \\ 0 & I_{\dnu}\end{bmatrix}}_{\Psi(w_{[1,i]})}\begin{bmatrix} \varphi_0\\ u(1)\\\vdots\\u(i-1)\end{bmatrix},
\end{equation}
\end{subequations}
where $n_{\Psi_i} = (n_\psi+1)(\dnu+n_{\Psi_{i-1}})$ and with $\theta_0 = \begin{bmatrix}a_{1} & a_2 & b_1 & b_2 \end{bmatrix}$, $\Psi(w_1) = \begin{bmatrix}\begin{bsmallmatrix} 1 \\ \psi(1) \end{bsmallmatrix}\kron I_{\dny} & 0 \\ 0 & \begin{bsmallmatrix} 1 \\ \psi(1) \end{bsmallmatrix}\kron I_{\dnu}\end{bmatrix}$ and $n_{\Psi_{0}}=\dny$ for this particular case. Note that the generalization for $n_\psi>1$ is trivial, as this only makes $a_2$ and $b_2$ row vectors, similarly for $\dnu>1$ and $\dny>1$. For $\dna,\dnb>1$, the definitions for $\Psi(w_0)$ and $n_{\Psi_0}$ change as multiple time-steps of $\psi(w_k)$ are involved in~\eqref{eq:nlvelioform-shifted}, which also implies that the recursive rule for the structure of $\theta_i$ slightly changes. 
\end{example}
This example shows that we can exploit a lot of structure in the formulation of a regressor-based multi-step predictor of the form~\eqref{eq:multisteppredictorformdesired}. 

\subsection{Deriving the model-based regressor form}
We now derive a regressor for IO representations of the form~\eqref{eq:nlvelioform}, {which also applies for~\eqref{eq:nlvelioform-shifted} and~\eqref{eq:nlvelioform-no-vel}}.  
The following observation, allows us to further characterize $\ms{T}(w_{[1,L]})$. 
The explored recursive formulation\footnote{The matrix function $\Psi$ in the shifted case will be dependent on $w_{[1-\dnr,i]}$ due to the definition of $w_k$ for~\eqref{eq:nlvelioform-shifted}. Compared to the general case, the matrix functions $\Psi(w_{[2,i]})$ will not change. Only $\Psi(w_{[1-\dnr,1]})$ must defined such that the shifts of the $\psi$'s are compatible with the initial trajectory.} can be decomposed in the individual time-steps in $w_{[1,i]}$ as follows:
\begin{multline}\label{eq:mtP}
    \Psi(w_{[1,i]}) = \begin{bmatrix} \begin{bsmallmatrix} 1 \\ \psi(w_i) \end{bsmallmatrix}\kron I_{n_{\Psi_i}} & 0 \\ 0 & \begin{bsmallmatrix} 1 \\ \psi(w_i) \end{bsmallmatrix}\kron I_{\dnu} \end{bmatrix} \begin{bmatrix} \begin{bsmallmatrix} 1 \\ \psi(w_{i-1}) \end{bsmallmatrix}\kron I_{n_{\Psi_{i-1}}} & 0 & 0 \\ 0 & \begin{bsmallmatrix} 1 \\ \psi(w_{i-1}) \end{bsmallmatrix}\kron I_{\dnu} & 0 \\ 0&0&I_{\dnu}\end{bmatrix} \times \cdots \\ \cdots \times 
    \begin{bmatrix} \begin{bsmallmatrix} 1 \\ \psi(w_{2}) \end{bsmallmatrix}\kron I_{n_{\Psi_{1}}} & 0 & 0 \\ 0 & \begin{bsmallmatrix} 1 \\ \psi(w_{2}) \end{bsmallmatrix}\kron I_{\dnu} & 0 \\ 0&0&I_{(i-1)\dnu}\end{bmatrix} \begin{bmatrix} \begin{bsmallmatrix} 1 \\ \psi(w_{1}) \end{bsmallmatrix}\kron I_{\dnr(\dnu+\dny)} & 0 \\ 0 & I_{i\dnu} \end{bmatrix}.
\end{multline}
Furthermore, because of the fact that the $(1,1)$-block of the first matrix of $\Psi(w_{[1,i]})$ is an identity matrix of size equal to the rows of $\Psi(w_{[1,i-1]})$, the specific time-relation w.r.t. $\psi(w_{i-1})$ for $\Delta y(i-1-j), \Delta u(i-1-j)$, $j=1,\dots,\dnr$ can be expressed by $\Psi(w_{[1,i]})$ using a sparse $\theta_i$. More importantly, we can write the multi-step predictor for~\eqref{eq:nlvelioform} as:
\begin{equation}\label{eq:predform}
    \Delta y_{[1,L]} = \Theta\,\Psi(w_{[1,L]})\begin{bmatrix} \Delta\varphi_0 \\ \Delta u_{[1,L]} \end{bmatrix}, 
\end{equation}
with $\Psi(w_{[1,L]})$ formulated as in~\eqref{eq:mtP} and
\begin{equation}\label{eq:defofTheta}
    \Theta := \begin{bmatrix} \begin{bmatrix}\theta_1 & 0 & \cdots & 0 \end{bmatrix} \\ \begin{bmatrix}\theta_2 & \cdots & 0 \end{bmatrix} \\ \vdots \\ \begin{bmatrix}\theta_{L-1} & 0 \end{bmatrix} \\ \theta_L \end{bmatrix},
\end{equation}
such that $\ms{T}(w_{[1,L]})= \Theta\,\Psi(w_{[1,L]})$ in~\eqref{eq:multisteppredictorformdesired}. Note that the multi-step predictors for the forms~\eqref{eq:nlvelioform-shifted} and~\eqref{eq:nlvelioform-no-vel} also satisfy this particular structure. Given the structure in the multi-step predictor, we can continue with the formulation of our \emph{kernelized} predictor.

\section{Kernelized data-driven representations}\label{s:kernel}
In this section, we derive a data-driven representation of primal and velocity forms in a kernelized form. In~\cite{huang2023robust} a kernelized form of a multi-step ahead predictor for general NL systems has  already been derived. This result has shown to be quite powerful, as illustrated in~\cite[Sec.~V]{huang2023robust}. However, it does not distinguishes the basis functions representing the coefficient function dependencies in~\eqref{eq:basisexpansion:a} and~\eqref{eq:basisexpansion:b}. Making it impossible to (i) increase the prediction length without a re-estimation step, or (ii) provide an efficient characterization of the NL-IO representation~\eqref{eq:io:gen} directly.

While the results of~\cite{huang2023robust} could readily be applied on the velocity form as well, the velocity form has a lot of structure that one should respect/exploit.  Furthermore, as we will show it later, an appropriate formulation of the kernel method is needed to distinguish the individual coefficient dependencies of~\eqref{eq:basisexpansion:a} and~\eqref{eq:basisexpansion:b} to use the recently developed LPV methods~\cite{verhoek2023direct, verhoek2023dpcjournal, Verhoek2023_dissipativity} or basis function approximation-based methods~\cite{lazar2024basis} for the analysis and control of NL systems. Our contribution is the highly non-trivial form of the RKHS estimator under the structural dependencies of the velocity form.

To obtain a kernelized multi-step predictor for \eqref{eq:nlvelioform}, we first introduce some preliminaries and important notions, followed by the presentation of the \emph{unstructured} kernelized predictor from~\cite{huang2023robust}. We conclude with the main result, i.e., the derivation of our structured predictor.

\subsection{Preliminaries and definitions}
We build upon the \emph{Reproducing Kernel Hilbert Space} (RHKS) framework to ensure flexibility and well-posedness of the problems we solve.
\begin{definition}
    An \emph{Reproducing Kernel Hilbert Space} (RHKS) over a non-empty set $\mb{W}$ is a Hilbert space of functions $\psi:\mb{W}\to\mb{R}$, such that for each $w\in\mb{W}$, $f(w)$ is bounded.
\end{definition}
A RKHS is associated with a positive semi-definite \emph{reproducing kernel}, which characterizes the inner product associated with the RKHS:
\begin{definition}\label{def:kernel}
    A symmetric function $\kappa:\mb{W}\times\mb{W}\to\mb{R}$ is called a positive semi-definite kernel if for any $h\in\mb{N}$
    \begin{equation}
        \sum_{i=1}^h\sum_{j=1}^h\alpha_i\alpha_j\kappa(w_i,w_j)\geq0, \quad \forall(w_k,\alpha_k)\in(\mb{W},\mb{R}), \ k=1,\dots,h.
    \end{equation}
    The \emph{kernel slice} of $\kappa$ centered at $\bar{w}$ is denoted $\kappa_{\bar{w}}(\cdot)=\kappa(\bar w, \cdot), \forall \bar{w}\in\mb{W}$. 
\end{definition}
The celebrated representer theorem states that an RKHS is fully characterized by its reproducing kernel. This gives that if a function $\psi:\mb{W}\to\mb{R}$ belongs to the RHKS $\meu{H}$ with kernel $\kappa$, then $\psi(w) = \lim_{h\to\infty}\sum_{i=1}^h \alpha_i\kappa_{{w}_i}(w)$. Furthermore, the inner product associated with the RKHS is expressed in terms of the kernel $\kappa$ as follows:
\[ \langle\psi,\phi\rangle = \lim_{h,h^\prime}\sum_{i=1}^h\sum_{j=1}^{h^\prime}\alpha_i\beta_j\kappa(w_i,w_j), \quad \text{where }\psi,\phi\in\meu{H}, \ \phi(w) = \lim_{h^\prime\to\infty}\sum_{i=1}^{h^\prime} \beta_i\kappa_{{w}_i}(w). \]
We will use kernels to construct a structured RHKS estimator of the velocity form using only data. Hence, suppose that we collected data from our data-generating NL system~\eqref{eq:io:gen} in the \emph{data-dictionary} $\mc{D}_N:= (\breve{u}(k), \breve{y}(k))_{k=0}^N$, where the breve-notation indicates measured data. We process the data by constructing the $\Delta$-signals and collect everything in Hankel matrices, which we conveniently write as:
\begin{equation} \label{eq:datamatrices}
    Y_\ell = \mc{H}_\ell(\Delta \breve{y}_{[1,N-L]}), \ Y_L = \mc{H}_L(\Delta \breve{y}_{[1+\ell,N]}), \
    U_\ell = \mc{H}_\ell(\Delta \breve{u}_{[1,N-L]}), \ U_L = \mc{H}_L(\Delta \breve{u}_{[1+\ell,N]}), \ W_{L} = \mc{H}_{L}(\breve{w}_{[1+\ell,N]}),
\end{equation}
where $\Delta\breve{y}_k = \breve{y}_k-\breve{y}_{k-1}$, similarly for $\Delta\breve{u}$, and $\breve{w}_k = \col\big(\breve{y}(k-1), \dots, \breve{y}(k-\ell-1), \breve{u}(k), \dots, \breve{u}(k-\ell-1)\big)$, cf.~\eqref{eq:defwk}.
Note that the width of these Hankel matrices, i.e., the number of columns, are equivalent, i.e., they all have $N-L-\ell+1$ columns. Let $N_\mr{c} = N-L-\ell+1$. The depth of the Hankel matrices $\ell$ and $L$ correspond to the length of the initial trajectory ($\ell\geq\dnr$) and the length of the predicted trajectory ($L\geq1$), respectively.

We are now ready to formulate the predictor for~\eqref{eq:nlvelioform}. We first recap the unstructured predictor that has been developed in~\cite{huang2023robust}, followed by our proposed \emph{structured} multi-step predictor for the velocity form.

\subsection{Unstructured kernelized multi-step predictor for \eqref{eq:nlvelioform}}\label{ss:unstructured}
In \cite{huang2023robust}, the following kernelized multi-step predictor is derived for nonlinear systems of the form~\eqref{eq:io:gen}:
\begin{equation}\label{eq:huang}
    y_{[1,L]} = \bar Y_L\big( \mbf{K} + \tfrac{1}{\gamma} I\big)^{-1} \begin{bmatrix} \kappa_{1}\big(\col(y_{[1-\ell,0]}, u_{[1-\ell,L]})\big) \\ \vdots \\ \kappa_{N_\mr{c}}\big(\col(y_{[1-\ell,0]}, u_{[1-\ell,L]})\big) \end{bmatrix},
\end{equation}
where $\bar Y_L=\mc{H}_L(\breve{y}_{[1+\ell,N]})$ and $\kappa_{i}\big(\col(y_{[1-\ell,0]}, u_{[1-\ell,L]})\big)= \kappa\big(\col(\breve{y}_{[i,i+\ell-1]},\breve{u}_{[i,i+\ell+L-1]}),\col(y_{[1-\ell,0]}, u_{[1-\ell,L]})\big)$. The matrix~$\mbf{K}$ is called the \emph{Gram} matrix, which is a positive semi-definite matrix of size $N_\mr{c}$ where the $(i,j)$\tss{th} element is constructed as 
\( \mbf{K}_{[i,j]} = \kappa\big(\col(\breve{y}_{[i,i+\ell-1]},\breve{u}_{[i,i+\ell+L-1]}),\col(\breve{y}_{[j,j+\ell-1]},\breve{u}_{[j,j+\ell+L-1]})\big).\)
This allows to write it in the implicit form
\begin{equation}
    \left[\begin{array}{c}
        \mbf{K} + \tfrac{1}{\gamma} I \\\hdashline \bar Y_L
    \end{array}\right]g=\left[\begin{array}{c} \kappa_{1}\big(\col(y_{[1-\ell,0]}, u_{[1-\ell,L]})\big) \\ \vdots \\ \kappa_{N_\mr{c}}\big(\col(y_{[1-\ell,0]}, u_{[1-\ell,L]})\big) \\\hdashline  y_{[1,L]}
    \end{array}\right],
\end{equation}
which is considered as the data-driven representation of the finite-horizon behavior of~\eqref{eq:io:gen}. 

This is a powerful result, and can be used to show equivalence with the (variations of) Willems' Fundamental Lemma~\cite{WillemsRapisardaMarkovskyMoor2005} in the exact case for special forms of~\eqref{eq:io:gen}~\cite{molodchyk2024exploring}, e.g., linear, Hammerstein or differentially flat systems. By writing~\eqref{eq:nlvelioform} as
\[ \Delta y(k) = f(\Delta y(k-1), \dots, \Delta y(k-\dna), \Delta u(k), \dots, \Delta u(k-\dnb), w_k), \]
we see that the result of~\cite{huang2023robust} can readily be applied to formulate a multi-step predictor and a data-driven representation of~\eqref{eq:nlvelioform}. However, in this case all the structure that is inherently present in~\eqref{eq:nlvelioform} is completely neglected. This will lead to a highly inefficient estimator, tedious kernel selection, and computationally heavy hyperparameter tuning, which requires much more data that even further increases the computational load. Hence, there is a need for a \emph{structured} form, which we will derive in the next section.

\subsection{Structured kernelized multi-step predictor for \eqref{eq:nlvelioform}}

We approach the problem of formulating a structured kernelized predictor for~\eqref{eq:nlvelioform} via the problem of estimating a ``model'' of the structured multi-step predictor~\eqref{eq:predform} using the data-dictionary $\mc{D}_N$. To establish this, we consider the following parametrized model structure
\begin{equation}\label{eq:regrmult2}
    \Delta \hat{y}_{[1,L]} = \Lambda\,\Psi(w_{[1, L]}) \begin{bmatrix} \Delta {\varphi}_0 \\ \Delta u_{[1,L]} \end{bmatrix} + e_{[1,L]}, \quad \text{and}\quad \Lambda = \begin{bmatrix} \lambda_1^\top \\ \vdots \\\lambda_{\dny L}^\top \end{bmatrix},
\end{equation}
where $e_{[1,L]}\in\mb{R}^{L\dny}$ is residual signal (the error of the fit) and $\Delta \hat{y}_{[1,L]}$ the model estimate. Here $\Psi$, according to the structure laid down in~\eqref{eq:mtP}, is based on the collection of unknown (possibly infinite number of) basis functions $\{\psi_i\}$. The parameter matrix $\Lambda$ is composed of parameter vectors $\lambda_i$ that are associated to this collection of basis functions (see also~\eqref{eq:basisexpansion}). The aim is to collectively estimate these from data. If the collection of basis functions in $\Psi$ coincides with the expansion in~\eqref{eq:basisexpansion}, the optimal solution for $\Lambda$ will coincide with $\Theta$.

In the estimation, we aim to minimize the error~$e$, while avoiding overfitting by regularizing the model parameters collected in $\Lambda$, given the data-set $\mc{D}_N$ measured from the NL system~\eqref{eq:io:gen}. 
\subsubsection{Ridge regression}
For every column of the Hankel matrices in~\eqref{eq:datamatrices}, the multi-step predictor~\eqref{eq:predform} can be written in terms of the data as:
\[ Y_{L_{[:,i]}} = \Theta\,\Psi(W_{L_{[:,i]}}) \begin{bmatrix} Y_{\ell_{[:,i]}} \\ U_{\ell_{[:,i]}} \\ U_{L_{[:,i]}} \end{bmatrix}, \quad i = 1, \dots, N_\mr{c}.  \]
where $Y_{\ell_{[:,i]}}$ indicates the $i$\tss{th} column of~$Y_\ell$.
Let us now construct the optimization problem that we want to solve for the estimation of~\eqref{eq:regrmult2}: 
\begin{align}
    \min_{\Lambda,e} \quad & \mc{J}(\Lambda, e) = \frac{1}{2}\|\Lambda\|^2_{2,2} + \frac{\gamma}{2}\sum_{k=1}^{N_\mr{c}}\left\|e_{[k,L+k-1]}\right\|^2_2 \\
    \mathrm{subject\ to} \quad &  e_{[k,L+k-1]} = Y_{L_{[:,k]}}-\Lambda\,\Psi(W_{L_{[:,k]}}) \begin{bmatrix} Y_{\ell_{[:,k]}} \\ U_{\ell_{[:,k]}} \\ U_{L_{[:,k]}} \end{bmatrix}
\end{align}
This constrained optimization problem is solved by constructing the Lagrangian:
\begin{equation}
    \mc{L}(\Lambda, e, \alpha) = \mc{J}(\Lambda, e) - \sum_{k=1}^{N_\mr{c}}\alpha_k\left(e_{[k,L+k-1]}- Y_{L_{[:,k]}}+\Lambda\,\Psi(W_{L_{[:,k]}}) \begin{bmatrix} Y_{\ell_{[:,k]}} \\ U_{\ell_{[:,k]}} \\ U_{L_{[:,k]}} \end{bmatrix}\right),
\end{equation}
with $\alpha_k\in\mb{R}^{1\times L\dny}$ the Lagrange multipliers. The optimum is obtained when the KKT conditions are satisfied, i.e., %
\begin{align}
    \frac{\partial \mc{L}}{\partial e} = 0 \to \quad & \quad \gamma e_{[k,L+k-1]} = \alpha_k^\top, \quad k = 1, \dots, N_\mr{c},\\
    \frac{\partial \mc{L}}{\partial \Lambda} = 0 \to \quad & \quad \Lambda = \sum_{k=1}^{N_\mr{c}}\alpha_k^\top \begin{bmatrix} Y_{\ell_{[:,k]}} \\ U_{\ell_{[:,k]}} \\ U_{L_{[:,k]}} \end{bmatrix}^\top \Psi^\top(W_{L_{[:,k]}}), \label{eq:KKT:Lambda}\\
    \frac{\partial \mc{L}}{\partial\alpha_i} = 0 \to \quad & \quad e_{[i,L+i-1]} = Y_{L_{[:,i]}}-\Lambda\,\Psi(W_{L_{[:,i]}}) \begin{bmatrix} Y_{\ell_{[:,i]}} \\ U_{\ell_{[:,i]}} \\ U_{L_{[:,i]}} \end{bmatrix}, \quad i = 1, \dots, N_\mr{c}.
\end{align}
Substitution of the former two in the latter gives
\begin{equation}
    Y_{L_{[:,i]}} = \left(\sum_{k=1}^{N_\mr{c}}\alpha_k^\top \begin{bmatrix} Y_{\ell_{[:,k]}} \\ U_{\ell_{[:,k]}} \\ U_{L_{[:,k]}} \end{bmatrix}^\top \Psi^\top(W_{L_{[:,k]}})\right) \Psi(W_{L_{[:,i]}}) \begin{bmatrix} Y_{\ell_{[:,i]}} \\ U_{\ell_{[:,i]}} \\ U_{L_{[:,i]}} \end{bmatrix} + \tfrac{1}{\gamma}\alpha_i^\top,
\end{equation}
which can be simplified to
\begin{equation}\label{eq:resultofsubstitutionKKT}
    Y_L = \mathrm{A}\left(\frac{1}{\gamma} I + \begin{bmatrix}
        X_{[:,1]} & 0 & \cdots & 0 \\ 0 & X_{[:,2]} & \ddots & \vdots \\ \vdots & \ddots & \ddots & 0 \\ 0 & \cdots & 0 & X_{[:,N_\mr{c}]} \end{bmatrix}^\top  \begin{bmatrix} \tilde{\Psi}_{1,1} & \tilde{\Psi}_{1,2} & \cdots & \tilde{\Psi}_{1,N_\mr{c}} \\ \tilde{\Psi}_{2,1} & \tilde{\Psi}_{2,2} & \ddots & \vdots \\ \vdots & \ddots & \ddots & \vdots \\ \tilde{\Psi}_{N_\mr{c},1} & \cdots & \cdots & \tilde{\Psi}_{N_\mr{c},N_\mr{c}} \end{bmatrix} \begin{bmatrix}
        X_{[:,1]} & 0 & \cdots & 0 \\ 0 & X_{[:,2]} & \ddots & \vdots \\ \vdots & \ddots & \ddots & 0 \\ 0 & \cdots & 0 & X_{[:,N_\mr{c}]} \end{bmatrix} \right),
\end{equation}
with $\mr{A} = \begin{bmatrix} \alpha_1^\top & \cdots & \alpha_{N_\mr{c}}^\top \end{bmatrix}$, $X_{[:,i]}=\begin{bmatrix} Y_{\ell_{[:,i]}} \\ U_{\ell_{[:,i]}} \\ U_{L_{[:,i]}} \end{bmatrix}$, and $\tilde{\Psi}_{i,j} = \Psi^\top(W_{L_{[:,i]}}) \Psi(W_{L_{[:,j]}})$. We already see a lot of structure appearing in~\eqref{eq:resultofsubstitutionKKT}. We can go further by exploring the structure of the individual matrix elements $\tilde{\Psi}_{i,j}$ in the middle block-matrix, using our findings in Section~\ref{s:regressor}.

\subsubsection{Structure of $\Psi^\top(W_{L_{[:,i]}}) \Psi(W_{L_{[:,j]}})$}
Let us first write out an element $\tilde{\Psi}_{i,j}$ of the middle block-matrix in~\eqref{eq:resultofsubstitutionKKT} with the structure derived in~\eqref{eq:mtP}:
\begin{multline}
    \Psi^\top(W_{L_{[:,i]}}) \Psi(W_{L_{[:,j]}}) = \begin{bmatrix} \begin{bsmallmatrix} 1 \\ \psi(\breve{w}_{i}) \end{bsmallmatrix}^\top\kron I_{(\dny+\dnu)\ell} & 0 \\ 0 & I_{L\dnu} \end{bmatrix} \times \dots \times \begin{bmatrix} \begin{bsmallmatrix} 1 \\ \psi(\breve{w}_{i+L-1}) \end{bsmallmatrix}^\top\kron I_{n_{\Psi_{L}}} & 0  \\ 0 & \begin{bsmallmatrix} 1 \\ \psi(\breve{w}_{i+L-1}) \end{bsmallmatrix}^\top\kron I_{\dnu}\end{bmatrix}  \\ \times \begin{bmatrix} \begin{bsmallmatrix} 1 \\ \psi(\breve{w}_{j+L-1}) \end{bsmallmatrix}\kron I_{n_{\Psi_{L}}} & 0  \\ 0 & \begin{bsmallmatrix} 1 \\ \psi(\breve{w}_{j+L-1}) \end{bsmallmatrix}\kron I_{\dnu}\end{bmatrix} \times \dots \times \begin{bmatrix} \begin{bsmallmatrix} 1 \\ \psi(\breve{w}_{j}) \end{bsmallmatrix}\kron I_{(\dny+\dnu)\ell} & 0 \\ 0 & I_{L\dnu} \end{bmatrix}.
\end{multline}
For the multiplication of the two matrices in the middle, we get elements on the diagonals of the form:
\begin{align*}
    \big(\begin{bsmallmatrix} 1 \\ \psi(\breve w_{i+L-1}) \end{bsmallmatrix}^\top\kron I_{\bullet}\big)\big(\begin{bsmallmatrix} 1 \\ \psi(\breve w_{j+L-1}) \end{bsmallmatrix}\kron I_{\bullet}\big) & = \left(\begin{bsmallmatrix} 1 & \psi(\breve w_{i+L-1})^\top \end{bsmallmatrix}\begin{bsmallmatrix} 1 \\ \psi(\breve w_{j+L-1}) \end{bsmallmatrix}\right)\kron I_\bullet = (1 + \psi(\breve w_{i+L-1})^\top\psi(\breve w_{j+L-1}))\kron I_\bullet \\ 
    & = \underbrace{(1+\psi(\breve w_{i+L-1})^\top\psi(\breve w_{j+L-1}))}_{\in\mb{R}} I_\bullet,
\end{align*}
where the second equality results from the Kronecker identity~\cite{hornjohnson91}. 
Taking out the scalar $(1+\psi(\breve w_{i+L-1})^\top\psi(\breve w_{j+L-1}))$ results in:
\begin{multline}
    \Psi^\top(W_{L_{[:,i]}}) \Psi(W_{L_{[:,j]}}) = (1+\psi(\breve w_{i+L-1})^\top\psi(\breve w_{j+L-1})) \begin{bmatrix} Q & 0 \\ 0 &  I_{\dnu} \end{bmatrix}, \quad\text{where} \\ 
    Q = \begin{bmatrix} \begin{bsmallmatrix} 1 \\ \psi(\breve{w}_{i}) \end{bsmallmatrix}^\top\kron I_{(\dnu+\dny)\ell} & 0 \\ 0 & I_{(L-1)\dnu} \end{bmatrix} \times \dots \times \begin{bmatrix} \begin{bsmallmatrix} 1 \\ \psi(\breve w_{i+L-2}) \end{bsmallmatrix}^\top\kron I_{n_{\Psi_{L-1}}} & 0  \\ 0 & \begin{bsmallmatrix} 1 \\ \psi(\breve w_{i+L-2}) \end{bsmallmatrix}^\top\kron I_{\dnu} \end{bmatrix}  \\ \times \begin{bmatrix} \begin{bsmallmatrix} 1 \\ \psi(\breve w_{j+L-2}) \end{bsmallmatrix}\kron I_{n_{\Psi_{L-1}}} & 0  \\ 0 & \begin{bsmallmatrix} 1 \\ \psi(\breve w_{j+L-2}) \end{bsmallmatrix}\kron I_{\dnu} \\ \end{bmatrix} \times \dots \times \begin{bmatrix} \begin{bsmallmatrix} 1 \\ \psi(\breve{w}_{j}) \end{bsmallmatrix}\kron I_{(\dnu+\dny)\ell} & 0 \\ 0 & I_{\dnu} \end{bmatrix},
\end{multline}
We can now apply the same trick recursively on the inner two matrices $L$ times, resulting in:
\begin{multline}\label{eq:structureKernel}
    \Psi^\top(W_{L_{[:,i]}}) \Psi(W_{L_{[:,j]}}) = \\ 
    \begin{bmatrix} 
        \prod_{t=0}^{L-1}(1+\psi(\breve w_{i+t})^\top\psi(\breve w_{j+t}))I_{(\dnu+\dny)\ell} & 0 & \cdots & 0 \\ 
        0 & \prod_{t=1}^{L-1} (1+\psi(\breve w_{i+t})^\top\psi(\breve w_{j+t}))I_{\dnu} & & \vdots \\
        \vdots  & & \ddots & 0 \\ 0 & \cdots \qquad \qquad \qquad \qquad \qquad 0 &  & (1+\psi(\breve w_{i+L-1})^\top\psi(\breve w_{j+L-1})) I_{\dnu}\end{bmatrix}.
\end{multline}
Hence, the elements $\tilde{\Psi}_{i,j}$ are constructed from multiplications of the \emph{inner products} $\langle\psi(\breve{w}_i),\psi(\breve{w}_j)\rangle$, $i,j=1,\dots,L$. We now apply the structured kernelization method by applying the kernel function $\kappa(\breve w_{i+L-1},\breve w_{j+L-1})$ for $\psi(\breve w_{i+L-1})^\top\psi(\breve w_{j+L-1})$, where $\kappa$ satisfies Definition~\ref{def:kernel}. Note that by the operations for kernel construction from kernels~\cite{cristianini2000introduction}, $\tilde{\kappa}(x,y):=1+\kappa(x,y)$ satisfies Definition~\ref{def:kernel} if $\kappa$ satisfies Definition~\ref{def:kernel}. Moreover, if $\kappa_1,\kappa_2$ satisfy Definition~\ref{def:kernel}, then $\kappa_3(x,y):=\kappa_1(x,y),\kappa_2(x,y)$ satisfies Definition~\ref{def:kernel}. This allows us to choose a positive definite kernel $\kappa(\cdot,\cdot)$ that defines the inner product of $\psi(\breve{w}_i),\psi(\breve{w}_j)$. Consequently, via the aforementioned kernel construction methods, $\kappa$ defines $\Psi^\top(W_{L_{[:,i]}}) \Psi(W_{L_{[:,j]}})$ such that we can write each element of $\tilde{\Psi}_{i,j}$ as a \emph{structured} kernel function $\mc{K}: \mb{R}^{L\dnw}\times\mb{R}^{L\dnw}\to\mb{R}^{\ell\dny+(\ell+L)\dnu\times\ell\dny+(\ell+L)\dnu}$ in accordance to~\eqref{eq:structureKernel}. Let %
\begin{equation}\label{eq:defKij}
    \mc{K}(W_{L_{[:,i]}},W_{L_{[:,j]}}) = \mr{blkdiag}\Big(\prod_{t=0}^{L-1}(1+\kappa(\breve w_{i+t},\breve w_{j+t})I_{(\dnu+\dny)\ell}, \prod_{t=1}^{L-1} (1+\kappa(\breve w_{i+t},\breve w_{j+t})I_{\dnu}, \dots, (1+\kappa(\breve w_{i+L-1},\breve w_{j+L-1}) I_{\dnu}  \Big),
\end{equation}
and introduce $K_{i,j}=\mc{K}(W_{L_{[:,i]}},W_{L_{[:,j]}})$, which is the evaluation of the kernel function $\mc{K}$ on the given data.
We can now write the inner block-matrix $\begin{bsmallmatrix} \tilde{\Psi}_{1,1} & \cdots & \tilde{\Psi}_{1,N_\mr{c}} \\ \svdots & \sddots & \svdots \\ \tilde{\Psi}_{N_\mr{c},1} & \cdots  & \tilde{\Psi}_{N_\mr{c},N_\mr{c}} \end{bsmallmatrix}$ in~\eqref{eq:resultofsubstitutionKKT}, which is dependent on the basis functions, as the block-matrix $\begin{bsmallmatrix} K_{1,1} & \cdots & K_{1,N_\mr{c}} \\ \svdots & \sddots & \svdots \\ K_{N_\mr{c},1} & \cdots  & K_{N_\mr{c},N_\mr{c}} \end{bsmallmatrix}$, which is dependent on the kernel functions. We refer to this matrix as the Gram matrix and is \emph{inherently} characterizing the basis function expansions of the matrix functions $\mf{a}_i,\mf{b}_j$ in~\eqref{eq:basisexpansion} due to the structured construction of the kernels. In the remainder, we derive an explicit and implicit form of the structured multi-step predictor~\eqref{eq:predform}, such that the implicit form may serve as a data-driven representation of the behavior of the velocity form. This is followed by a discussion on the possibilities and open questions that will be addressed later.

\subsubsection{Formulation of the structured explicit and implicit predictors}
We now formulate a predictor based on only the data (in terms of the Hankel matrices), the Gram matrix and the so-called \emph{kernel slices}.
Define the following matrices:
\begin{equation}
    \mbf{X}_{N_\mr{c}} = \begin{bmatrix}X_{[:,1]} & 0 & \cdots & 0 \\ 0 & X_{[:,2]} & \ddots & \vdots \\ \vdots & \ddots & \ddots & 0 \\ 0 & \cdots & 0 & X_{[:,N_\mr{c}]} \end{bmatrix},\qquad
    \mbf{K}_{N_\mr{c}} = \begin{bmatrix} K_{1,1} & K_{1,2} & \cdots & K_{1,N_\mr{c}} \\ K_{2,1} & K_{2,2} & \ddots & \vdots \\ \vdots & \ddots & \ddots & \vdots \\ K_{N_\mr{c},1} & \cdots & \cdots & K_{N_\mr{c},N_\mr{c}} \end{bmatrix},
\end{equation}
with $K_{i,j}$ as in~\eqref{eq:defKij}.
From~\eqref{eq:resultofsubstitutionKKT}, we see that the alpha-matrix $\mr{A}$ is computed as:
\begin{equation}\label{eq:defofAlpha}
    \mr{A} = Y_L\left(\tfrac{1}{\gamma}I + \mbf{X}_{N_\mr{c}}^\top\mbf{K}_{N_\mr{c}}\mbf{X}_{N_\mr{c}}\right)^{-1}.
\end{equation}
Now consider~\eqref{eq:KKT:Lambda}, which can be rewritten as:
\begin{equation}\label{eq:KKT:Lambda:rewritten}
    \Lambda = \mr{A}\begin{bmatrix} X_{[:,1]}^\top \Psi^\top(W_{L_{[:,1]}}) \\ \vdots \\ X_{[:,N_\mr{c}]}^\top \Psi^\top(W_{L_{[:,N_\mr{c}]}}) \end{bmatrix}.
\end{equation}
Substitution of~\eqref{eq:defofAlpha} into~\eqref{eq:KKT:Lambda:rewritten} gives:
\begin{equation}
    \Lambda = Y_L\left(\tfrac{1}{\gamma}I + \mbf{X}_{N_\mr{c}}^\top\mbf{K}_{N_\mr{c}}\mbf{X}_{N_\mr{c}}\right)^{-1}\begin{bmatrix} X_{[:,1]}^\top \Psi^\top(W_{L_{[:,1]}}) \\ \vdots \\ X_{[:,N_\mr{c}]}^\top \Psi^\top(W_{L_{[:,N_\mr{c}]}}) \end{bmatrix},
\end{equation}
which we, in turn, can substitute in our considered model structure~\eqref{eq:regrmult2}:
\begin{subequations}\label{eq:explicitpredictor}
\begin{align}
    \Delta \hat{y}_{[1,L]} & = \left(Y_L\left(\tfrac{1}{\gamma}I + \mbf{X}_{N_\mr{c}}^\top\mbf{K}_{N_\mr{c}}\mbf{X}_{N_\mr{c}}\right)^{-1}\begin{bmatrix} X_{[:,1]}^\top \Psi^\top(W_{L_{[:,1]}}) \\ \vdots \\ X_{[:,N_\mr{c}]}^\top \Psi^\top(W_{L_{[:,N_\mr{c}]}}) \end{bmatrix}\right)\Psi(w_{[1, L]}) \begin{bmatrix} \Delta {\varphi}_0 \\ \Delta u_{[1,L]} \end{bmatrix}  \\
    & =  Y_L\left(\tfrac{1}{\gamma}I + \mbf{X}_{N_\mr{c}}^\top\mbf{K}_{N_\mr{c}}\mbf{X}_{N_\mr{c}}\right)^{-1} \mbf{X}_{N_\mr{c}}^\top\begin{bmatrix} \Psi^\top(W_{L_{[:,1]}})\Psi(w_{[1, L]}) \\ \vdots \\ \Psi^\top(W_{L_{[:,N_\mr{c}]}})\Psi(w_{[1, L]}) \end{bmatrix}\begin{bmatrix} \Delta {\varphi}_0 \\ \Delta u_{[1,L]} \end{bmatrix}, \\
    & = Y_L\left(\tfrac{1}{\gamma}I + \mbf{X}_{N_\mr{c}}^\top\mbf{K}_{N_\mr{c}}\mbf{X}_{N_\mr{c}}\right)^{-1} \mbf{X}_{N_\mr{c}}^\top\begin{bmatrix} K_1(w_{[1, L]}) \\ \vdots \\ K_{N_\mr{c}}(w_{[1, L]}) \end{bmatrix}\begin{bmatrix} \Delta {\varphi}_0 \\ \Delta u_{[1,L]} \end{bmatrix},
\end{align}
\end{subequations}
with the kernel slices $K_i(w_{[1, L]})= \Psi^\top(W_{L_{[:,i]}}) \Psi(w_{[1, L]})=\mc{K}(W_{L_{[:,i]}},w_{[1, L]})$, which have the same structure as in~\eqref{eq:defKij}. We have now obtained our multistep predictor, i.e., given a choice for the kernel, data matrices~\eqref{eq:datamatrices} and an initial trajectory and input trajectory, we can solve for the output $\Delta\hat y_{[1, L]}$ corresponding to that particular initial trajectory and the given input.
Note that this development can be seen as a structured version of the explicit predictors presented in Section~\ref{ss:unstructured}. {In fact, the result of~\cite{huang2023robust} is recovered if we disregard the structure completely, i.e., if we would choose some kernel $\kappa(\cdot,\cdot)$, e.g., a Radial Basis Function (RBF) kernel, and simply choose
\begin{equation}\label{eq:huangcase}
    X_{[:,i]}^\top \Psi^\top(W_{L_{[:,i]}}) \Psi(W_{L_{[:,j]}}) X_{[:,j]}: = \kappa\left(\begin{bsmallmatrix}X_{[:,i]} \\ W_{L_{[:,i]}} \end{bsmallmatrix} ,\begin{bsmallmatrix}X_{[:,j]} \\ W_{L_{[:,j]}} \end{bsmallmatrix}\right)
\end{equation}
we would arrive at~\eqref{eq:huang}, which would be an inefficient estimator given the rich structure of the underlying functional dependency. 

Before we derive the implicit formulation, we want to highlight that our form~\eqref{eq:explicitpredictor} can be easily scaled back and forth for various choices of $\ell$ and $L$. Moreover, it is even possible to explicitly compute the matrix coefficient functions $\mf{a}_i,\mf{b}_j$ through the characterization~\eqref{eq:basisexpansion} with a different substitution routine of the KKT conditions to obtain an explicit data-driven model of the velocity form.}

From the explicit structured predictor~\eqref{eq:explicitpredictor}, we formulate the implicit multi-step predictor that may serve as the data-driven finite-horizon representation of the velocity form: 
\begin{multline}
    g := \left(\tfrac{1}{\gamma}I + \mbf{X}_{N_\mr{c}}^\top\mbf{K}_{N_\mr{c}}\mbf{X}_{N_\mr{c}}\right)^{-1} \mbf{X}_{N_\mr{c}}^\top\begin{bmatrix} K_1(w_{[1, L]}) \\ \vdots \\ K_{N_\mr{c}}(w_{[1, L]}) \end{bmatrix}\begin{bmatrix} \Delta {\varphi}_0 \\ \Delta u_{[1,L]} \end{bmatrix} \\\ \iff \left(\tfrac{1}{\gamma}I + \mbf{X}_{N_\mr{c}}^\top\mbf{K}_{N_\mr{c}}\mbf{X}_{N_\mr{c}}\right)g = \mbf{X}_{N_\mr{c}}^\top\begin{bmatrix} K_1(w_{[1, L]}) \\ \vdots \\ K_{N_\mr{c}}(w_{[1, L]}) \end{bmatrix}\begin{bmatrix} \Delta {\varphi}_0 \\ \Delta u_{[1,L]} \end{bmatrix},
\end{multline}
yielding the implicit representation:
\begin{equation}
    \begin{bmatrix}
        \tfrac{1}{\gamma}I + \mbf{X}_{N_\mr{c}}^\top\mbf{K}_{N_\mr{c}}\mbf{X}_{N_\mr{c}} \\ Y_L
    \end{bmatrix}g = \left[\begin{array}{c|c}
         \mbf{X}_{N_\mr{c}}^\top \begin{bmatrix} K_1(w_{[1, L]}) \\ \vdots \\ K_{N_\mr{c}}(w_{[1, L]}) \end{bmatrix} & I_{L\dny}
    \end{array}\right] \left[\begin{array}{c} \Delta \hat{\varphi}_0 \\ \Delta \hat{u}_{[1,L]} \\\hline\Delta \hat{y}_{[1,L]} \end{array}\right].
\end{equation}
This formulation can be considered to be the finite-horizon data-driven representation of the velocity form~\eqref{eq:nlvelioform} of the nonlinear system~\eqref{eq:io:gen}. Finally, we want to emphasize that these structured representations and predictors are readily applicable for (i)~\mbox{1-step}-ahead predictors, when choosing $L=1$, see also the state-feedback case in~\cite{verhoek2023direct}, (ii)~the special cases of the velocity form, see Section~\ref{ss:shiftvel}, Appendix~\ref{app:deriveshiftedvel}, and (iii)~the direct factorization of the primal form, given in Appendix~\ref{app:primalform}.

\section{Using the structured kernelized predictors}
We have now established methods to obtain multi-step predictors and data-driven representations of velocity forms of nonlinear systems. These methods are also applicable for primal forms of the nonlinear system. In this section, we give a brief overview on how to use these representations in analysis and control problems.

\subsection{Usage for data-driven analysis and control}
In recent years, the use of finite-horizon data-driven representations in data-driven analysis and control has been increasing rapidly, cf.,~\cite{dePersisTesi2020, coulson2019data}. The representation that we presented here is readily available in this context, see, e.g., the predictive control application of the unstructured case in~\cite{huang2023robust}. The main problem with the currently available data-driven methods is that they focus either on LTI systems, or (approximations of) the primal form of nonlinear systems. While the problem for the former is evident, the problem with the latter is more subtile. When performing (data-driven) analysis and control of the primal form, one will often only get \emph{local guarantees}, e.g., only performance is guaranteed in the neighborhood around the origin of the nonlinear system. As motivated in Section~\ref{s:intro}, one way\footnote{Over the years, there are various concepts developed in the field of nonlinear analysis and control that study this problem, such as contraction analysis, incremental analysis, equilibrium-independent analysis, convergence analysis, etc. We are convinced that the velocity-based analysis is the most useful for the data-driven setting.} to overcome this is by using the velocity form of the nonlinear system. This is because guarantees around the origin of the velocity form imply these guarantees around any arbitrary (forced) equilibrium point of the nonlinear system~\cite{koelewijn2024convex}. Hence, for velocity-based analysis, the local guarantees of the velocity form translate to global guarantees of the primal form.

\begin{figure}[!ht]
    \begin{center}
        \includegraphics[scale=1]{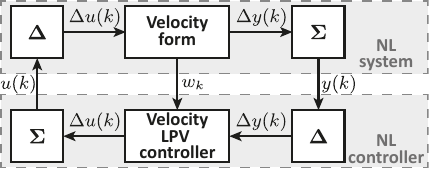}
        \caption{Possible realization for a controller that is designed for the velocity form.}\label{fig:control-realization}
    \end{center}
\end{figure}

For velocity-based controller synthesis using classical design method, one designs a controller that will stabilize the velocity-form. To ensure that the locally obtained guarantees translate to global guarantees for the closed-loop primal form, one will need a \emph{realization} of the controller `back' to the primal domain that preserves this. One way to establish this is via the sum $\mbf{\Sigma}$ and difference $\mbf{\Delta}$ operator, i.e., $\mbf{\Sigma}\Delta x_k = x_k$, $\mbf{\Delta}x_k = \Delta x_k$, and $\mbf{\Delta}(\mbf{\Sigma}\Delta x_k) = \Delta x_k$. With these operators, a possible realization of the controller is depicted in Fig.~\ref{fig:control-realization}, see also~\cite{Koelewijn2023, verhoek2023direct}.

One way to establish analysis and controller design with the presented representations is via the data-driven LPV framework. The connection between the NL-IO forms, our predictors and the LPV framework is discussed in the next section.

\subsection{Connecting the NL-IO forms and kernelized predictors to the LPV framework}
Given the discussed IO forms of the velocity representation of the NL system, the resulting quasi-linear structure can be also reformulated as a \emph{linear-parameter-varying} (LPV) representation {by means of the embedding principle}.
We end up with the LPV embedding of the velocity form of \eqref{eq:io:gen} if we now define the signal $p$ as the concatenation of the basis functions $\psi_\ell$ in~\eqref{eq:basisexpansion}, i.e.,
\begin{equation}\label{eq:p=basis}
    p(k) = \begin{bmatrix} \psi_1(w_k) \\ \vdots \\ \psi_{\dnp}(w_k) \end{bmatrix} \in\mb{P}\subseteq\mb{R}^{\dnp},  \quad \text{and} \quad \mb{P}\subseteq\psi(\mb{W})=\psi(\underbrace{\mb{Y}\times\dots\times\mb{Y}}_{\dna \text{ times}}\times\underbrace{\mb{U}\times\dots\times \mb{U}}_{\dnb+1 \text{ times}})
\end{equation}
we obtain the LPV IO representation:
\begin{equation}\label{eq:lpvioembstaticgen}
    \Delta y(k) = \sum_{i=1}^{\dna}a_i(p(k))\Delta y(k-i) + \sum_{j=1}^{\dnb}{b}_j(p(k))\Delta u(k-j),
\end{equation}
with $a_i(p(k)) = a_{i,0} + \sum_{\ell=1}^{\dnp} a_{i,\ell}p_\ell(k)$ and $b_j(p(k)) = b_{j,0} + \sum_{\ell=1}^{\dnp} b_{j,\ell}p_\ell(k)$. 
More importantly, if we consider the shifted form~\eqref{eq:nlvelioform-shifted} and we \emph{know} the basis functions, we can directly apply the LPV fundamental lemma~\cite{VerhoekTothHaesaertKoch2021} to obtain an exact data-driven representation of the velocity form. More precisely, if we know the basis function expansion~\eqref{eq:basisexpansion} of~\eqref{eq:nlvelioform-shifted},
we can define the signal $p$ as a concatenation of the basis functions:
\begin{equation*}
    p(k) = \begin{bmatrix} \psi_1(w_k) \\ \vdots \\ \psi_{\dnp}(w_k) \end{bmatrix} \in\mb{P}\subseteq\mb{R}^{\dnp}, \quad \text{and} \quad \mb{P}\subseteq\psi(\mb{Y},\mb{Y},\mb{U},\mb{U})
\end{equation*}
which results in the \emph{shifted-affine} LPV IO realization that serves as an embedding of the velocity form of~\eqref{eq:io:str1}:
\begin{equation}
    \Delta y(k) = \sum_{i=1}^{n}\left(\bar{a}_i\big(p(k-i)\big)\Delta y(k-i) + \bar{b}_i\big(p(k-i)\big)\Delta u(k-i)\right),
\end{equation}
with $\bar{a}_i(p(k-i)) = \bar{a}_{i,0} + \sum_{\ell=1}^{n} \bar{a}_{i,\ell}p_\ell(k-i)$ and $\bar{b}_i(p(k-i)) = \bar{b}_{i,0} + \sum_{\ell=1}^{n} \bar{b}_{i,\ell}p_\ell(k-i)$. 
Then, 
\[ \begin{bmatrix}
\mc{H}_{L}(\Delta \breve{u}_{[1,N]}) \\
\mc{H}_{L}(\Delta \breve{y}_{[1,N]}) \\
\mc{H}_{L}(\Delta \breve{u}^{\breve{\mt{p}}}_{[1,N]}) - \mc{P}^{\dnu}\mc{H}_{L}(\Delta\breve{u}_{[1,N]}) \\
\mc{H}_{L}(\Delta \breve{y}^{\breve{\mt{p}}}_{[1,N]}) - \mc{P}^{\dny}\mc{H}_{L}(\Delta\breve{y}_{[1,N]})
\end{bmatrix} g = \begin{bmatrix}
    \Delta u_{[1,L]} \\
    \Delta y_{[1,L]} \\
    0\\
    0
\end{bmatrix},  \]
with $\mc{P}^{\bullet}=\mathrm{blkdiag}(p(1)\kron I_\bullet, \dots, p(L)\kron I_\bullet)$ and $x^\mt{p}_{[1,N]}=(p(k)\kron x(k))_{k=1}^{N}$, serves as a \emph{direct} data-driven representation of the horizon-$L$ behavior of the velocity form embedding. With this representation, we can directly perform analysis and control using the existing tools~\cite{Verhoek2023_dissipativity,verhoek2023dpcjournal}.

\section{Example}

\begin{figure}[t]
    \centering
    \includegraphics[scale=0.7]{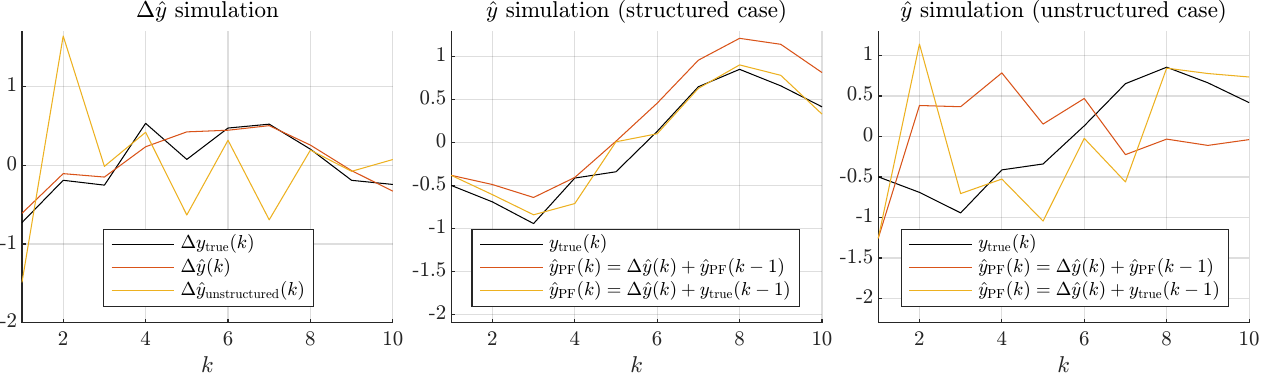}
    \caption{Simulation results for the example system.}\label{fig:technote-example}
\end{figure}

In this example, we demonstrate the effectiveness of the proposed data-driven representations that can be used for data-driven analysis and control of NL systems with global guarantees. We will discuss a simulation example using structured and unstructured kernelized predictors of the velocity form of the considered NL system. We consider the SISO NL example system:
\begin{equation}\label{eq:examplesys}
    y(k) = -u(k-2)e^{-y^2(k-1)}+0.5y(k-2)u^2(k-1).
\end{equation}
To excite the system and obtain our data dictionary, we feed an i.i.d. white noise input signal to the system with mean~$0$ and variance~$1$. We measure the output $\breve{y}$ of~\eqref{eq:examplesys} with a white noise output error, i.e., $\breve{y}(k) = y(k)+e(k)$, where $\mr{variance}(e) = 0.1$. The length of the data-dictionary is $900$, i.e., $N=899$. Given an input trajectory and initial trajectory, i.e., a regressor, the aim is to predict an output trajectory of $L=10$ steps. We choose $\ell=2$. In this example, we take the commonly chosen RBF kernel $\kappa(w_i, w_j) = \exp(-\|w_i - w_j\|_2^2/\sigma^2)$, with hyperparameter $\sigma$. We perform two simulation studies: (i)~Estimation of a structured predictor of the velocity form with the method discussed in this note, (ii)~Estimation of an unstructured predictor of the velocity form via the methods in~\cite{huang2023robust}, i.e.,~\eqref{eq:huangcase}. For both of the studies, we perform a grid-based hyperparameter optimization for the hyperparameters $\gamma$ and $\sigma$. For the first study, the hyperparameter optimization gave $\sigma=40.11$ and $\gamma=123.3$. The result for the second study provided $\sigma=25.97$ and $\gamma=1474.5$. If we simulate the resulting predictors of the velocity form, and compare the estimate $\Delta \hat{y}$ with the true output $\Delta {y}_\mr{true}$, we obtain the left plot in Fig.~\ref{fig:technote-example}. For this simulation, we assume we know the true $w_k$ (this assumption can be alleviated with, e.g., iterative estimation methods). We see that the structured multi-step predictor predicts the `true' trajectory of the velocity form much better than the unstructured one. Moreover, if we estimate the trajectory of the primal form using the prediction of the velocity form and the previous time-step of the (a) predicted output, i.e., $y_\mr{PF}(k) = \Delta \hat{y}(k) + \hat{y}_\mr{PF}(k)$, or (b) the true output, i.e., $y_\mr{PF}(k) = \Delta \hat{y}(k) + {y}_\mr{true}(k)$, we obtain the right two plots in Fig.~\ref{fig:technote-example}. Here, the middle plot is the result with our structured predictor, while the right plot is the result with the unstructured predictor of~\cite{huang2023robust}. We again see that the structured predictor estimates the output of the primal form much better than the unstructured predictor. We have tried to improve the unstructured results by additionally performing the study with the following extended kernels
\begin{align*}
    \kappa(w_i, w_j) & = w_i^\top w_j\big(1+\exp(-\|w_i - w_j\|_2^2/\sigma^2)\big),\\
    \kappa(w_i, w_j) & = (1+w_i^\top w_j)\exp(-\|w_i - w_j\|_2^2/\sigma^2), \\
    \kappa(w_i, w_j) & = w_i^\top w_j+ \exp(-\|w_i - w_j\|_2^2/\sigma^2), \\
    \kappa(w_i, w_j) &= w_i^\top w_j \exp(-\|w_i - w_j\|_2^2/\sigma^2).
\end{align*}
These, unfortunately, did not improve the results any further. On the other hand, this shows the power of the structure exploitation in our proposed kernelized predictors.

\section{Conclusions and future work}

In this technical note, we established the formulation of structured kernel-based data-driven representations of the velocity form. These predictors can be used to achieve data-driven analysis and control of nonlinear systems with \emph{global} stability and performance guarantees. The predictors we formulated are highly flexible and are applicable to both the velocity form and the primal form of the nonlinear system. We have showed the advantage of the structured predictor compared to an unstructured one in a simulation example. Future work involves a priori enforcing properties of the predictor through proper kernel selection and application of the structured predictor in data-driven analysis and control.

\appendix
\subsection{Using the FTC for {recasting primal} NL-IO forms}\label{app:primalform}

As discussed in~\cite{KoelewijnToth2021AutomaticEmbedding}, we can also write general NL-IO forms as~\eqref{eq:io:gen}, i.e. the primal form, as an IO form that is quasi-linear in the signal $y(k-i)$ and $u(k-i)$. For this we need to take the additional assumption that $f(0)=0$. Then, defining
\[ w_k := \begin{bmatrix} y(k-1) &  \cdots & y(k-\dna) & u(k) & \cdots &  u(k-\dnb) \end{bmatrix}, \]
and considering the `wrapper' function $\bar{f}(\lambda):= f(\lambda w_k)$ with $\lambda\in[0,1]$, we know with the FTC that
\[ \bar{f}(1)-\bar{f}(0) = \int_0^1\frac{\partial \bar{f}(\bar\lambda)}{\partial\lambda}\mr{d}\bar\lambda, \] 
which results in
\begin{align}
    {f}(w_k)-\underbrace{\bar{f}(0)}_{0} & = \int_0^1\frac{\partial {f}(\bar\lambda w_k)}{\partial\zeta}w_k\mr{d}\bar\lambda, \notag\\
    y(k) = f(w_k) & = \left(\int_0^1\frac{\partial {f}(\bar\lambda w_k)}{\partial\zeta}\mr{d}\bar\lambda\right)w_k, \notag \\
    & = \sum_{i=1}^{\dna}\underbrace{\left(\int_0^1\frac{\partial {f}(\bar\lambda w_k)}{\partial y(k-i)}\mr{d}\bar\lambda\right)}_{\hat{\mf{a}}_i(w_k)}y(k-i) + \sum_{i=0}^{\dnb}\underbrace{\left(\int_0^1\frac{\partial {f}(\bar\lambda w_k)}{\partial u(k-i)}\mr{d}\bar\lambda\right)}_{\hat{\mf{b}}_j(w_k)}u(k-i).\label{eq:nlvelioform-no-vel}
\end{align}
Note however that, despite this form \emph{looks} linear in $y(k-i)$ and $u(k-j)$, it is in-fact \emph{not}! This is because the matrix functions $\hat{\mf{a}}_i,\hat{\mf{b}}_j$ are \emph{also} dependent on $y(k-i)$, $i=1,\dots,\dna$  and $u(k-j)$, $j=0,\dots,\dnb$. {Similar as with the velocity form, assuming a shifted NL-IO representation simplifies the dependency of the coefficient matrix functions, resulting in $\hat{\mf{a}}_i(w_{k-i})$ and $\hat{\mf{b}}_j(w_{k-j})$.}

\subsection{Derivation of~\eqref{eq:nlvelioform-shifted} from~\eqref{eq:io:str1}}\label{app:deriveshiftedvel}
We start at~\eqref{eq:io:str1}. As in Section~\ref{ss:genvel}, we again construct the time-difference form by analyzing $(y(k)-y(k-1),u(k)-u(k-1))=:(\Delta y(k),\Delta u(k))$. This yields:
\begin{align*}
    y(k)-y(k-1) & = \sum_{i=1}^{n} f_i(y(k-i),u(k-i)) - \sum_{i=1}^{n} f_i(y(k-i-1),u(k-i-1))  \\ 
    & = f_1(y(k-1),u(k-1)) - f_1(y(k-2),u(k-2)) + \cdots \notag \\ 
    & \hspace{2cm} \cdots + f_n(y(k-n),u(k-n)) - f_n(y(k-n-1),u(k-n-1)).
\end{align*}
Applying the FTC gives:
\begin{multline*}
    \Delta y(k) = \left(\int_0^1\frac{\partial f_1}{\partial y}(\bar{y}(k-1,\lambda), \bar{u}(k-1,\lambda))\mr{d}\lambda\right)\Delta y(k-1) +  \\
    + \left(\int_0^1\frac{\partial f_1}{\partial u}(\bar{y}(k-1,\lambda), \bar{u}(k-1,\lambda))\mr{d}\lambda\right)\Delta u(k-1) + \cdots \\
    \cdots + \left(\int_0^1\frac{\partial f_n}{\partial y}(\bar{y}(k-n,\lambda), \bar{u}(k-n,\lambda))\mr{d}\lambda\right)\Delta y(k-n) + \\
    + \left(\int_0^1\frac{\partial f_n}{\partial u}(\bar{y}(k-n,\lambda), \bar{u}(k-n,\lambda))\mr{d}\lambda\right)\Delta u(k-n),
\end{multline*}
where 
\begin{align*}
    \bar{y}(k-i,\lambda) & = y(k-i-1) + \lambda(y(k-i)-y(k-i-1)), \quad \text{and} \\
    \bar{u}(k-i,\lambda) & = u(k-i-1) + \lambda(u(k-i)-u(k-i-1)).
\end{align*}
We can now write this compactly by defining a matrix function for every integral, e.g.,
\begin{align*}
    \bar{\mf{a}}_1(y(k-1),y(k-2), u(k-1),u(k-2))&:= \int_0^1\frac{\partial f_1}{\partial y}(\bar{y}(k-1,\lambda), \bar{u}(k-1,\lambda))\mr{d}\lambda, \\ \vdots \hspace{3cm} & \hspace{3cm} \vdots \\
    \bar{\mf{a}}_n(y(k-n),y(k-n-1), u(k-n),u(k-n-1))&:= \int_0^1\frac{\partial f_n}{\partial y}(\bar{y}(k-n,\lambda), \bar{u}(k-n,\lambda))\mr{d}\lambda, \\
    \bar{\mf{b}}_1(y(k-1),y(k-2), u(k-1),u(k-2))&:= \int_0^1\frac{\partial f_1}{\partial u}(\bar{y}(k-1,\lambda), \bar{u}(k-1,\lambda))\mr{d}\lambda, \\ \vdots \hspace{3cm} & \hspace{3cm}\vdots\\
    \bar{\mf{b}}_n(y(k-n),y(k-n-1), u(k-n),u(k-n-1))&:= \int_0^1\frac{\partial f_n}{\partial u}(\bar{y}(k-n,\lambda), \bar{u}(k-n,\lambda))\mr{d}\lambda.
\end{align*}
With $w_k:= \mr{col}(y(k),y(k-1),u(k),u(k-1))$, this yields the compact IO representation~\eqref{eq:nlvelioform-shifted}, which is the velocity form of~\eqref{eq:io:str1}.

\bibliographystyle{IEEEtran}
\bibliography{ref_nonlinear}
\end{document}